\documentclass[apjl]{emulateapj}
\usepackage{graphicx, graphics}

\received{}
\revised{}
\accepted{}

\shorttitle{Hot Gas in the Milky Way Halo}
\shortauthors{Fang, Bullock, \& Boylan-Kolchin}

\begin{document}

\title{On the Hot Gas Content of the Milky Way Halo}

\author{Taotao~Fang\altaffilmark{1,2}, James Bullock\altaffilmark{2}, and
  Michael Boylan-Kolchin\altaffilmark{2,3}}

\altaffiltext{1}{Department of Astronomy and Institute of Theoretical Physics and
  Astrophysics, Xiamen University, Xiamen, Fujian 361005, China} 
\altaffiltext{2}{Department of Physics \& Astronomy, 4129 Frederick Reines Hall,
  University of California, Irvine, CA 92697} 
\altaffiltext{3}{Center for Galaxy Evolution Fellow}
\begin{abstract}
  The Milky Way appears to be missing baryons, as the observed mass in stars and
  gas is well below the cosmic mean. One possibility is that a substantial
  fraction of the Galaxy's baryons are embedded within an extended,
  million-degree hot halo, an idea supported indirectly by observations of warm
  gas clouds in the halo and gas-free dwarf spheroidal satellites.  X-ray
  observations have established that hot gas does exist in our Galaxy beyond the
  local hot bubble; however, it may be distributed in a hot disk configuration.
  Moreover, recent investigations into the X-ray constraints have suggested that
  any Galactic corona must be insignificant.  Here we re-examine the
  observational data, particularly in the X-ray and radio bands, in order to
  determine whether it is possible for a substantial fraction of the Galaxy's
  baryons to exist in $\sim 10^6$ K gas.  In agreement with past studies, we
  find that a baryonically closed halo is clearly ruled out if one assumes that
  the hot corona is distributed with a cuspy NFW profile.  However, if the hot
  corona of the galaxy is in an extended, low-density distribution with a large
  central core, as expected for an adiabatic gas in hydrostatic equilibrium,
  then it may contain up to $10^{11}$ M$_{\odot}$ of material, possibly
  accounting for all of the missing Galactic baryons.  We briefly discuss some
  potential avenues for discriminating between a massive, extended hot halo and
  a local hot disk.
\end{abstract}


\section{Introduction}
\label{sec:intro}
The question of whether or not significant reservoirs of hot baryons exist
around bright field galaxies remains a topic of current debate
(\citealp{dai2010,anderson2010,humphrey2011,prochaska2011}) despite years of
discussion in the literature (\citealp{spitzer1956, bahcall1969, mo1996,mo2002}).
Theoretical prejudice favors the idea that quasi-stable $\sim 10^6$ K coronae
develop as shock-heated material in the aftermath of halo collapse
\citep{silk1977, white1978, white1991, benson2000, keres2005, dekel2006,
  keres2009, crain2010}, though the density structure and mass content of these
halos is sensitive to uncertain physics and energy injection processes
(\citealp{white1978, maller2004, kaufmann2009, benson2010, sharma2012}).

If hot gaseous halos exist, they would provide a potential hiding place for {\em
  missing galactic baryons} -- the $\gtrsim 80\%$ of baryons that are
unaccounted for by collapsed gas and stars in galaxies (e.g.,
\citealp{fukugita1998, anderson2010, behroozi2010, moster2012}).  It is not
known whether the missing halo baryons exist primarily as diffuse hot gas halos
around normal galaxies (\citealp{maller2004, fukugita2006, sommer-larsen2006}),
have been mostly expelled as a result of energetic blow out (\citealp{dekel1986,
  almeida2008}), or were never accreted in the first place, possibly as a result
of pre-heating (e.g. \citealp{mo2005}). So far, searches have failed to detect
extended X-ray emission around nearby spiral galaxies (e.g.,
\citealp{bregman1998, benson2000, rasmussen2009a}). Most of the detected X-ray
emission in these galaxies is centered around disk/bulge region, and is
associated with active star formation regions (see, e.g., \citealp{wang2007,
  li2007}).

Locally, the Milky Way provides an important benchmark for understanding the
missing galactic baryon problem.  The Galaxy's dark matter halo mass is somewhat
uncertain, but maser observations, stellar halo tracers, and satellite
kinematics suggest a total virial mass\footnote{We define the virial mass using
  an average enclosed density of $\Delta = 95$ times the critical density of the
  Universe, as determined using the spherical top-hat collapse model in the LCDM
  cosmology.} in the range $M_{\rm v} = (1-2) \times 10^{12}$ M$_\odot$ (see
\citealt{boylan-kolchin2012} for a summary), within an associated virial radius
$R_{\rm v} \simeq (260 - 330)$ kpc.  In the absence of mass loss, the Milky
Way's baryonic allotment therefore should be $M_{\rm b} = f_{\rm b} \, M_{\rm v}
\simeq (1.65 - 3.3) \times 10^{11}$ M$_\odot$, assuming a universal baryon
fraction $f_b = 0.165$ (consistent with \citealp{komatsu2011}).  The observed
cold baryonic mass of the Milky Way is well below this, $M_{\star} \simeq 0.65
\times 10^{11}$ \citet{mcmillan2012}, with effectively negligible contributions
from cool disk gas \citep{kalberla2009} and satellite galaxies
\citep{mcconnachie2012}.  At least $10^{11}$ M$_\odot$ of baryons are missing
from the Galactic census.

 
There is circumstantial evidence that at least some of these missing Galactic
baryons are in an extended, hot corona.  For example, a confining hot medium can
help explain gas clouds in the Magellanic Stream \citep{stanimirovic2002}, and
the shapes of shells along the edge of the Large Magellanic Cloud
\citep{de-boer1998} are similarly suggestive.  H-alpha emission at the leading
edge of the Magellanic stream is best understood via heating from a fairly dense
hot medium ($\sim 10^{-4}\rm\ cm^{-3}$ at $\sim 50$ kpc;
\citealp{weiner1996}). The thermal pressure of high-velocity clouds at the same
distance is consistent with a similar confining medium \citep{fox2005}.  More
recent ultraviolet and optical absorption probes of the Magellanic Stream
provide strong evidence for a complex, multi-phase structure that is being
evaporated by interactions with a hot corona \citep{fox2010}.  The head-tail
structure of high velocity clouds is another observation best explained in the
context of a confining corona \citep{putman2011}.  A separate argument for the
existence of such a medium is the lack of detected HI in most dwarf spheroidal
galaxies within 270 kpc of the Milky Way \citep{grcevich2009} -- a result that
is naturally explained by ram pressure stripping by a $\sim 10^{-4}\rm\ cm^{-3}$
hot medium at $\sim 70$ kpc (see also \citealp{moore1994, lin1983,
  nichols2011}).

While these indirect probes are enlightening, the most direct avenue for
detecting a hot corona is via X-ray studies. It is well known that the soft
X-ray background (between 0.1 and 2 keV) consists of three components: the Local
Hot Bubble (LHB), extragalactic emission (mostly from active galactic nuclei),
and a thermal component that lies at an intermediate distance (see, e.g.,
\citealp{kuntz2000}). One popular interpretation of this thermal component is
that it originates from the hot interstellar medium (ISM) in the Galactic disk,
i.e., a hot gas disk rather than a $\sim 100$ kpc extended Galactic halo.
Indeed, using joint X-ray emission-absorption analysis, \citet{yao2009} and
\citet{hagihara2010} argued that the hot gas in our Galaxy is confined within a
few kpc around the stellar disk.  The detection of the highly ionized metal
absorption lines at $z=0$ in numerous quasar spectra also suggested a hot gas
component in and around our Galaxy (see, e.g., \citealp{nicastro2002, fang2003,
  fang2006, bregman2007a, gupta2012}); however, it is also unclear this
absorption is produced by hot gas in the disk or distant
halo. \citet{anderson2010} focused specifically on the question of baryonic
closure, asking whether the dispersion measure of pulsars in the Large
Magellanic Cloud (LMC) could be reconciled with a baryonically closed Milky Way
halo.  
They concluded that it could not, though they focused mainly on cases where the hot
gas follows a cuspy density profile characterized by either a single power law
or a broken power law (as expected for dissipationless dark matter).

In what follows, we re-examine the question of the the Milky Way's hot corona in
the context of X-ray surface brightness constraints and pulsar dispersion
measurements from the standpoint of three illustrative models for the hot gas
density distribution: (1) an extended profile with a central core, as expected
for an adiabatic gas in hydrostatic equilibrium (\citealt{maller2004}, hereafter
MB); (2) a centrally concentrated \citet*[NFW]{navarro1997} profile
and (3) a local hot gas disk.  The first profile is among the puffiest
distributions one might consider, though it is in fact very similar to
predictions for realistic galaxy-size halos with gaseous halos that are in both
hydrostatic and thermal equilibrium \citep{sharma2012}.  The second profile we
consider, the NFW hot halo, is the most centrally concentrated distribution that
one could envision for a hot gas around a galaxy, given that it is the profile
that arises from the collapse of non-interacting dust.  While it is unclear how
such a profile could arise in a real astrophysical plasma, this is an assumption
that often appears in the literature.  The third case explores a scenario where
all of the X-ray gas detections are explained by a local distribution with
globally negligible mass content ($< 10^8$ M$_\odot$).

\section{Model Definitions}
\label{sec:models}
In our fiducial explorations, we assume that the Milky Way's dark matter halo
has a virial mass of $M_{\rm v} = 10^{12}\rm M_{\odot}$ (with associated radius
$R_{\rm v} = 260$ kpc) and that it follows an NFW profile with a concentration of
$C_{\rm v} = 12$ (and thus a scale radius $R_s = 21.7$ kpc; \citealp{navarro1997,
  bullock2001}).  The implied mass in missing baryons is approximately
$10^{11}\rm M_{\odot}$.  We then explore the three aforementioned hot gas
distributions, including two coronal models that contain $M_{\rm hot} = 10^{11}$
M$_{\odot}$ of material, in this context.

\subsection{Extended Adiabatic Halo: MB}
\label{subsec:mb}
\begin{figure*}[t]
\center
\includegraphics[height=.85\textwidth,angle=180]{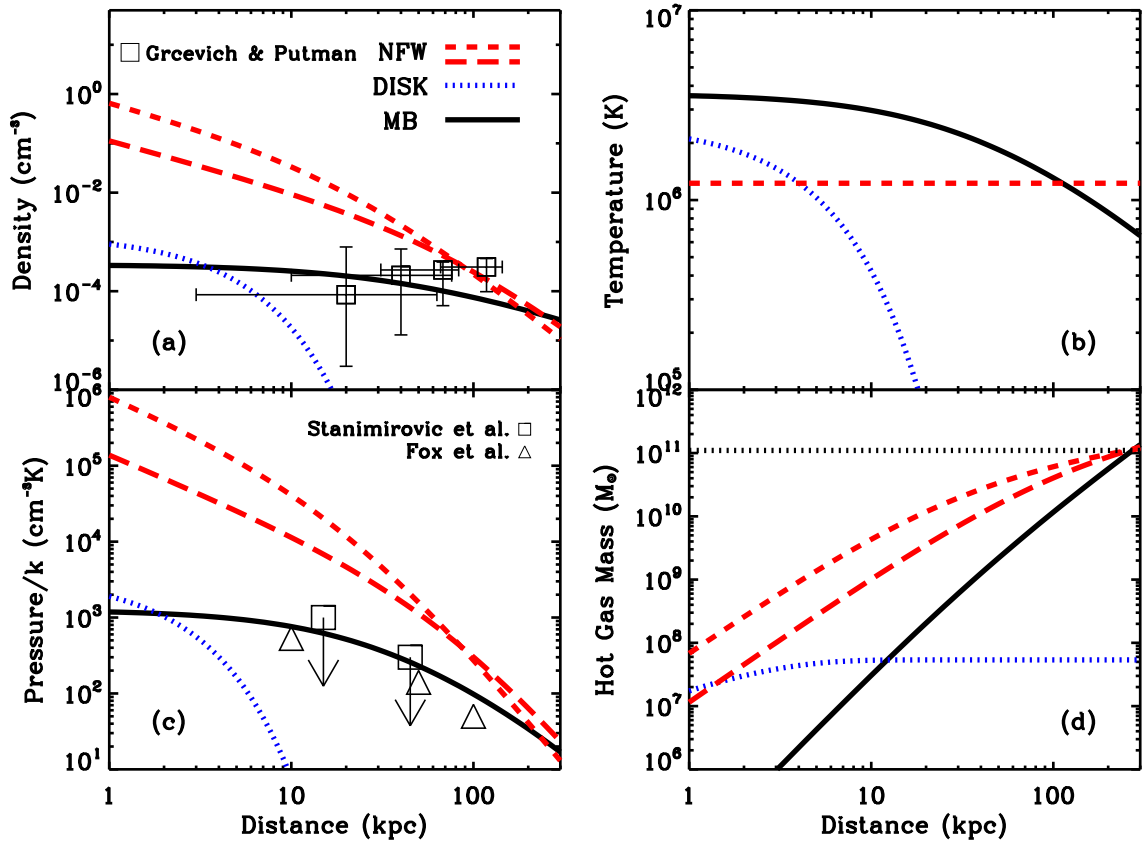}
\vskip-1cm
\caption{Hot gas profiles as a function of radius. \textit{Panel (a)}: density;
  \textit{panel (b)}: temperature; \textit{panel (c)}: pressure; \textit{panel
    (d)}: hot gas mass. The predictions from the MB model are plotted in solid
  black, the DISK model in dotted blue, and NFW models with $C_{\rm v}=12$ (3) in red
  dash (dot-dash). For the DISK model, the $x$-axis is actually the vertical
  distance ($z$) from the disk.  The square symbols in panel (a) show estimates
  of the hot gas density required to explain the lack of HI in Milky Way dwarf
  galaxies as derived from ram pressure striping arguments by
  \citet{grcevich2009}. The square symbols in panel (c) are derived by
  \citet{stanimirovic2002} and \citet{fox2005} under the assumption that high
  velocity clouds in the Milky Way halo are pressure-confined.  See text for
  details.}
\label{fig:profile}
\end{figure*}

The first model we explore, MB, assumes that the hot halo is distributed as
adiabatic gas with polytropic index of 5/3 that is in hydrostatic equilibrium
within the Milky Way's NFW dark matter halo of concentration $C_{\rm v}$ \citep{maller2004}:
\begin{equation}
\rho^{\rm MB}_g(r)=\rho_{\rm v} \left[1+\frac{3.7}{x}\ln(1+x)-\frac{3.7}{C_{\rm v}}\ln(1+C_{\rm v})\right]^{3/2},
\end{equation}
where $r$ is the radius from the Galactic center and $x\equiv r/R_s$, with $R_s$
the scale radius of the dark matter halo.  The density $\rho_{\rm v}$ is the gas
density at the virial radius.  In our fiducial case, we chose $\rho_{\rm v}$ such that
the integrated hot gas mass within $R_{\rm v}$ is $M_{\rm hot} = 10^{11} \rm
M_{\odot}$.  The resulting temperature profile is
\begin{equation}
T^{\rm MB}_g(r)=T_{\rm v} \left[1+\frac{3.7}{x}\ln(1+x)-\frac{3.7}{C_{\rm v}}\ln(1+C_{\rm v})\right],
\end{equation}
where $T_{\rm v}$ is the hot halo temperature at $R_{\rm v}$.  We set $T_{\rm v}$ by requiring
that the X-ray emission-weighted temperature be consistent with {\sl XMM}-Newton
and {\sl Suzaku} observations, which both give halo temperatures consistent with
$T_h = 0.2$ keV, with errors of $\sim$ 10\%. The normalization, $T_{\rm v}$, is then
determined by:
\begin{equation}
T_h = \frac{\int \rho^2(r)\Lambda(T,Z_g)T(r)dr}{\int \rho^2(r)\Lambda(T,Z_g)dr},
\end{equation}
where $\Lambda(T,Z_g)$ is the cooling function \citep{sutherland1993} and we
assume a $Z_g = 0.3 Z_{\odot}$ (results for different choices of $Z_g$ are
explored in Section~\ref{subsec:param_dependence}).  The implied density,
temperature, pressure, and mass profiles for the MB model are shown as solid
black lines in Figure~\ref{fig:profile}.

\subsection{Cuspy Halo: NFW}
\label{subsec:NFW}
As mentioned in the introduction, it is common to make the simplifying
assumption that the density distribution of the hot gas around galaxies traces
what is expected for the dark matter alone:
\begin{equation}
\rho^{\rm NFW}_g(r) = \frac{\rho_g}{x(1+x)^2}.
\end{equation}
As before, $x\equiv r/R_s$, where $R_s \equiv R_{\rm v}/C_h$ is the NFW scale radius.
We explicitly allow $C_h$ to be different than $C_{\rm v}$ of the background dark
matter halo.  As before, we fix the normalization $\rho_g$ by requiring that the
total hot gas mass within the virial radius equals $10^{11} \rm M_{\odot}$.  For
the NFW profile, we assume a constant hot halo temperature of $T_h$.  This
requires an unusual equation of state profile that varies with radius for
self-consistency, but we adopt it here for comparison to previous work.
  
We explore two cases for the NFW hot halo concentration: $C_h= 12$ (which mimics
the dark matter background exactly: $C_h = C_{\rm v}$) and also a low-concentration
case with $C_h = 3$.  The implied density, temperature, pressure, and mass
profiles for the high (low) concentration NFW case are shown by the dashed
(dot-dashed) lines in Figure~\ref{fig:profile}.

\subsection{Local Model: DISK}
\label{subsec:disk}
For comparison, we also calculate the hot gas distribution in an exponential
disk model, which has been favored by recent X-ray observations. Following \citet{yao2009a}, we adopt a vertical distribution
\begin{equation}
\rho^{\rm DISK}_g(z) = \rho_0\exp\left(-\frac{z}{h_{\rho}\xi}\right),
\end{equation}
and
\begin{equation}
T^{\rm DISK}_g(z) = T_0\exp\left(-\frac{z}{h_{T}\xi}\right),
\end{equation}
Here $\xi$ is the volume filling factor and is assumed to be 1, $z$ is the
vertical distance from the disk, $\rho_0$ and $T_0$ are the gas density and
temperature at the disk mid-plane, and $h_{\rho}$ and $h_T$ are the scale height
of the hot gas density and temperature distributions, respectively.

We set $\rho_0 = 1.4\times10^{-3}\rm\ cm^{-3}$, $h_{\rho} = 2.3$ kpc, $T_0 = 10^{6.4}$K, and $h_T = 5.6$ kpc based on
observations of the PKS~2155-304 sight line
\citep{hagihara2010}. \citet{yao2009} observed the LMC X-3 sight line and
reached similar conclusion. This model is plotted in blue in
Figure~\ref{fig:profile}; the $y$-axis is the vertical distance from the
Galactic disk ($|z|$) since in the model the density and temperature depends
only on this distance.

\section{Observational Constraints}
\label{sec:constraints}
\begin{figure*}[t]
\center
\centerline{\includegraphics[width=1.\textwidth,height=.3\textheight,angle=0]{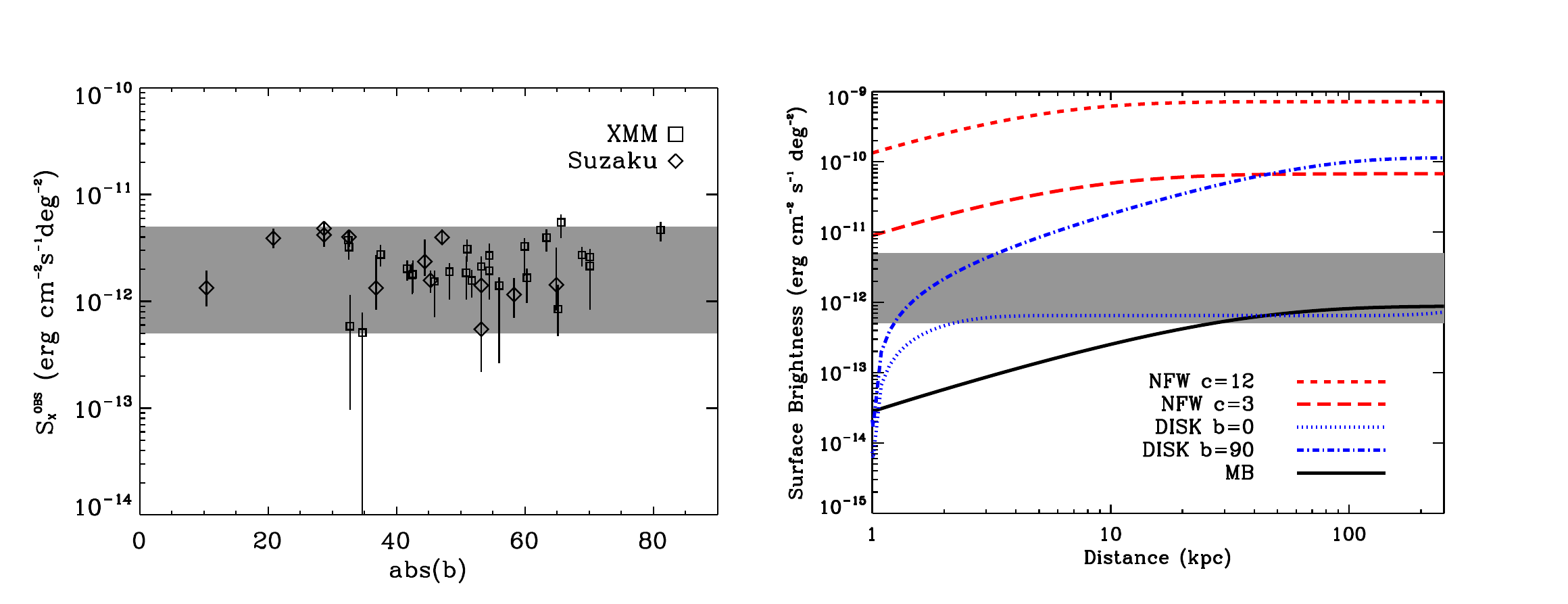}}
\caption{X-ray emission constraints.  \textit{Left panel}: X-ray surface
  brightness between 0.4 and 2 keV as a function of Galactic latitude. {\sl XMM}
  data are shown as squares, and {\sl Suzaku} as diamonds (see text for
  details). The grey region shows the range of variation. \textit{Right panel}:
  Model predictions of the X-ray surface brightness between 0.4 and 2 keV as a
  function of radius. Lines are the same as shown in Figure~\ref{fig:profile};
  however, we plot the DISK model prediction for two $b$ values, 0$^\circ$
  (upper blue dotted curve) and 90$^\circ$ (lower blue dotted curve). The grey
  area is reproduced from the left panel.}
\label{fig:xray}
\end{figure*}

\subsection{Indirect Probes}
\label{subsec:indirect_probes}
One of the strongest pieces of indirect evidence for an extended hot gas
reservoir around the Galaxy comes from the lack of gas detected in small dwarf
satellite galaxies around the Milky Way.  Dwarf galaxies tend to be gas rich
unless they are within close ($\sim 300$ kpc) proximity of a larger system, an
observation that is usually interpreted as arising from ram-pressure stripping
\citealp{lin1983, moore1994}.

Recently, \citet{grcevich2009} examined the HI content of the Local Group dwarf
galaxies. They found a cut-off radius of $\sim 270$ kpc around the Milky Way or
Andromeda, below which no dwarf spheroidal galaxy contains detected HI gas. They
argued the ram-pressure stripping produced by a hot, extended halo gas can
explain the lack of the HI gas in these dwarfs, and they constrained this gas
density by the measurements of four dwarfs, Carina, Ursa Minor, Sculptor, and
Fornax, between $\sim20$ to 100 kpc. The data points and their error bars are
plotted in the top left panel of Figure~\ref{fig:profile}.  These measurements
are much too extended to be explained by a local disk model (blue dotted) and
too low density to be explained by the cuspy NFW models (red); they are matched
reasonably well by the MB distribution, however.

A completely distinct indication for extended hot gas comes from studying gas
clouds around the galaxy.  Relying on the Arecibo telescope,
\citet{stanimirovic2002} observed the Magellanic Stream in HI 21-cm
emission. They found that the most likely mechanism to confine the stream clouds
is via pressure from a hot halo.  They placed upper limits on the hot gas
pressure at $10^3$ and $3\times 10^{-2}\rm\ cm^{-3}K$ at 15 and 45 kpc,
respectively (square symbols in Figure~\ref{fig:profile}, bottom left
panel). \citet{fox2005} have reached similar conclusions when studying high
velocity clouds (HVCs) in the halo.  These authors concluded that if HVCs are
pressure-confined, then the pressure of the surrounding medium must be
approximately 530, 140, and 50 $\rm\ cm^{-3}K$, at distances of 10, 50, and 100
kpc, respectively (triangles in Figure~\ref{fig:profile}, bottom left panel).
Again, while neither the NFW nor DISK models matches these constraints, the
extended MB model provides pressure support at approximately the required level.

Of course, neither the cloud-confinement arguments nor the ram-pressure
stripping arguments individually demands the existence of an extended hot halo.
It may be possible that some other environmental process can explain the lack of
gas in local dwarfs, though there is no obvious candidate for such a mechanism.
Similarly, it is possible that warm gas clouds in the Galactic halo are not
pressure-confined.  The confinement could, in principle, be gravitational if the
clouds are embedded within dark matter halos \citep{sternberg2002}.
Nevertheless, it is interesting to recognize that an extended hot gas halo
containing a substantial fraction of the halo's baryons is capable of matching
both constraints quite well.

\subsection{X-ray Emission}
\label{subsec:xray}
X-ray emission provides a strong constraint on the extent and density of hot gas
around the Galaxy.  We consider data from two recent surveys with the {\sl
  XMM}-Newton and {\sl Suzaku} X-ray telescopes. \citet{henley2010} studied 26
high-latitude {\sl XMM}-Newton observations of diffuse X-ray emission ($|b| >
30^{\circ}$ where $b$ is the Galactic latitude). The Galactic longitudes ($l$)
of these observations were selected to be between 120$^{\circ}$ and
240$^{\circ}$ to avoid enhanced emission from the Galactic center.
\citet{yoshino2009} studied the soft, diffuse X-ray emission in 12 fields
observed with the {\sl Suzaku} X-ray Telescope. The Galactic longitudes and
latitudes are similar to those {\sl XMM} fields, i.e., high latitudes and away
from the direction of the Galactic center, except two fields at $b=10^{\circ}$
and $20^{\circ}$.

While the exact details differ between the two analyses, the general procedures
are similar (see their papers for details). After cleaning the spectrum, they
fitted the observed spectrum to three components: (1) the cosmic X-ray
background; (2) local thermal plasma emission at a temperature around 0.1 keV
from solar wind charge exchange (SWCX) and local hot bubble (LHB) emission; and
(3) distant thermal emission from near the Galaxy but beyond the LHB. The third
component is the one that we consider here as likely arising from the distant
halo.

In the XMM data analysis, a single power law was adopted to model the cosmic
X-ray background, so we use the results presented in Table 2 of Henley et
al. For the {\sl Suzaku} data, various models for the Cosmic X-ray background
were tried and we present the the one that relied on a single power law (Yoshino
et al.'s Table 4) in order to directly compare with the XMM result.  We select
the energy band between 0.4 and 2 keV (the default value adopted in the XMM data
analysis) to compare the observed and modeled X-ray surface brightness ($S^{\rm
  OBS}_X$). Since for the {\sl Suzaku} data $S^{\rm OBS}_X$ were not calculated,
we estimate the $S^{\rm OBS}_X$ using their models with the software package
XSPEC; specifically, we use their model 1' (see their Table~4), which adopted a
power law model of the CXB.  The resultant normalized data is plotted in the
left panel of Figure~\ref{fig:xray}.  The grey area indicates the full range of
the data, which we will use as a benchmark comparison for our model predictions.

We make predictions for the X-ray surface brightness from each of our hot gas
models using
\begin{equation}
S_X = \frac{1}{4\pi}\int\rho_e(r)\rho_i(r)\Lambda[T(r)]dr.
\end{equation}
Here $\rho_e$ and $\rho_i$ are the electron and proton densities, and
$\Lambda(T)$ is the cooling function. In order to compare with observations, we
use the Astrophysical Plasma Emission Code (APEC)\footnote{See
  http://atomdb.org/.} to calculate the X-ray emissivity between 0.4 and 2 keV
as a function of plasma temperature. Since the X-ray emissivity is also a
function of metal abundance, we assume a $Z_g = 0.3 Z_{\odot}$ for the MB and
NFW models (\citealp{cen2006a, rasmussen2009}), and 1 $Z_{\odot}$ for
the DISK model, where $Z_{\odot}$ is the solar abundance of \citet{anders1989}.  We explore the effect of these choices on our predictions below.

In the right panel of Figure~\ref{fig:xray}, we plot the predicted surface
brightness for our three models as the integrated emission coming from gas
within a given distance (horizontal axis). The grey band is the constraint from
observations -- the upper limits of the X-ray emission from the hot diffuse
gas. Note that even though both {\sl XMM}-Newton and {\sl Suzaku} observations avoid the
X-ray bright sky toward the Galactic center, we expect some sightlines may still
experience local enhancement from supernova heating.

Figure~\ref{fig:xray} clearly indicates the NFW profiles predict far too much
X-ray emission.  In particular, the $C_{\rm v} = 12$ case exceeds observations by more
than two orders of magnitude. This is consistent with the results of
\citet{anderson2010}, and it comes about because the density at the center of
the NFW profile is far too high (see the left panel of
Figure~\ref{fig:profile}).  Indeed, the $C_{\rm v}=3$ profile also substantially
over-predicts the X-ray emission. We see from the rise of the line with distance
that the majority of the X-ray emission comes from the inner $\sim$ 10 kpc for
both of the NFW models (red lines). Since the $C_h = 12$ hot gas NFW model is
ruled out to an extreme degree, we will only include the $C_h =3$ (low
concentration) NFW halo in our comparisons for the remainder of this paper.

In contrast to the NFW model, the MB profile (solid black) is fully consistent
with the data, comfortably reaching the lower part of the observed region at
$\sim 25\,{\rm kpc}$. Recall that the MB profile has the same mass in hot
baryons within $R_{\rm v}$ as the NFW models, so the spatial distribution of the gas
obviously plays a crucial role in determining the X-ray properties of the halo.
The two dotted blue curves in Figure~\ref{fig:xray} represent the the DISK model
predictions for in-plane ($b=0^{\circ}$; upper curve) and vertical
($b=90^{\circ}$; lower curve) sight lines. The DISK model can also fit the
observed value quite well at high Galactic latitude; however, at low Galactic
latitude, it also predicts too much X-ray emission.  This is partially a limitation of the simplified disk model assumed, which does
not truncate in the radial plane.  Unfortunately, with the
current data we cannot tell whether there is a increasing trend at low $b$ (see
the right panel of Figure~\ref{fig:profile}); however, future targeted
observations at low $b$ may help distinguish between an extended hot halo like
MB and a local hot distribution like the DISK model.

\subsection{Pulsar Dispersion Measure}
\label{subsec:pulsars}
\begin{figure*}[!t]
\center
\includegraphics[width=.49\textwidth,height=.3\textheight,angle=180]{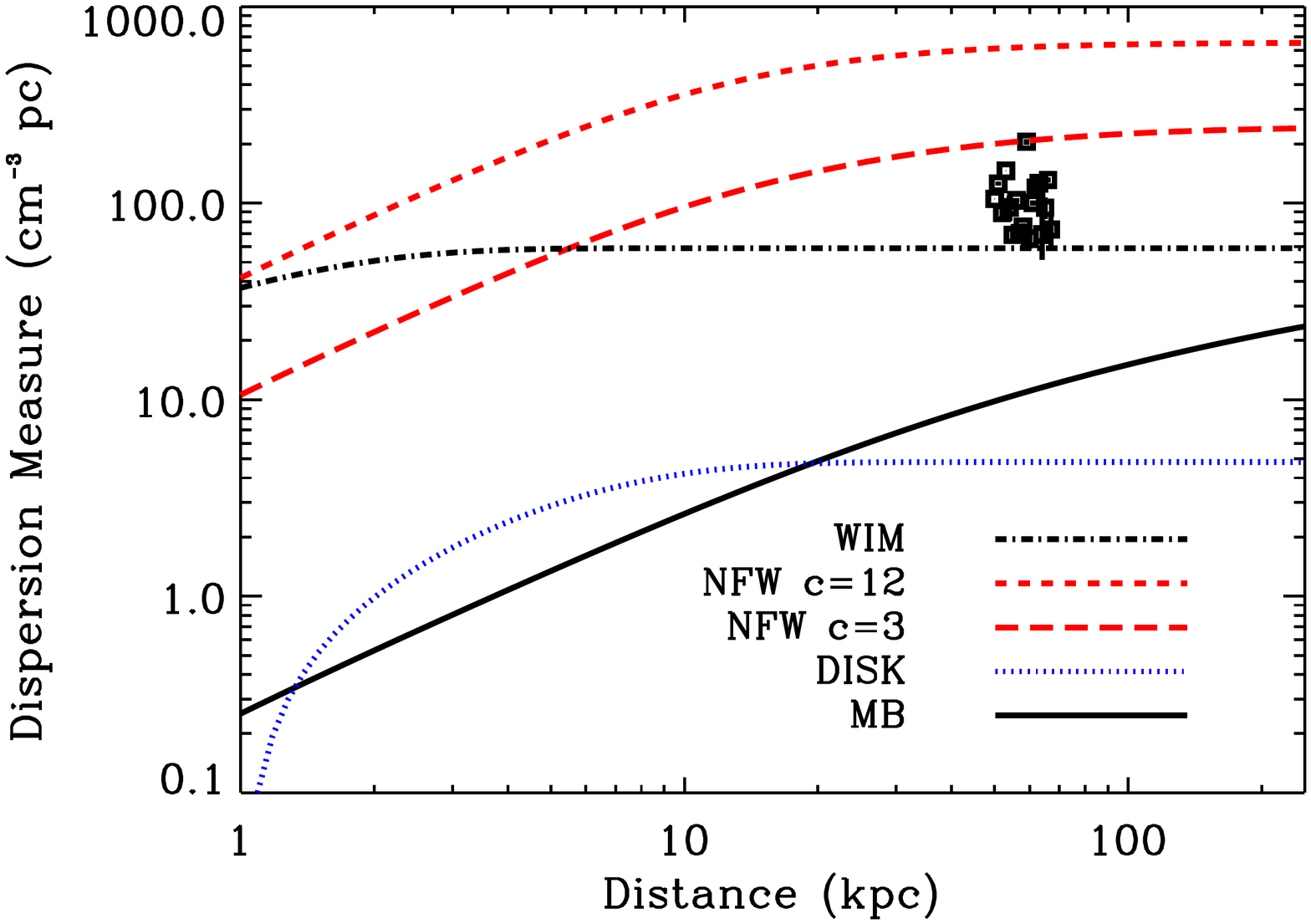}
\includegraphics[width=.49\textwidth,height=.3\textheight,angle=180]{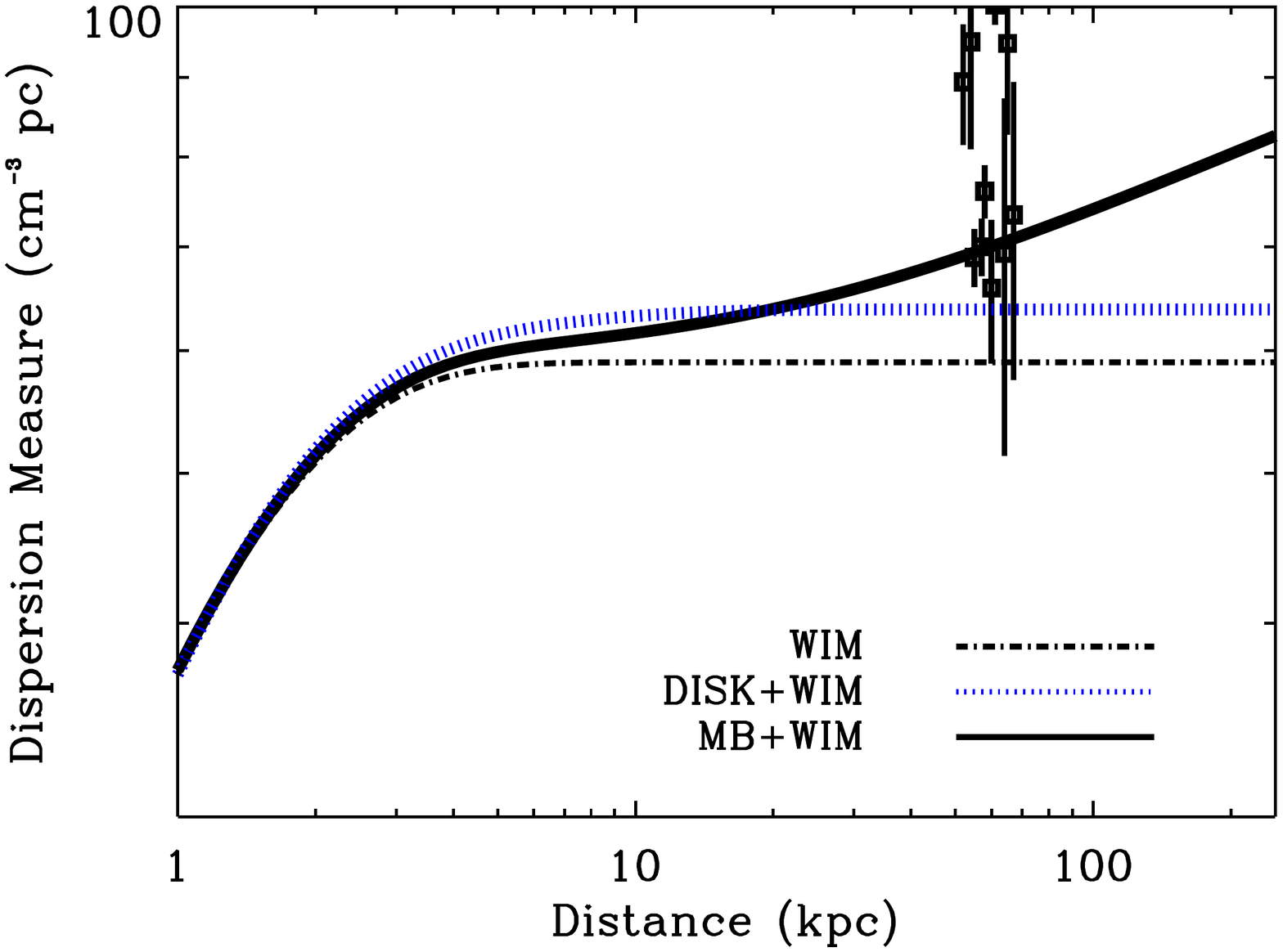}
\caption{\textit{Left panel}: Pulsar dispersion measure as a function of distance. The dash-dotted line shows
  the contribution from the Warm Ionized Medium (WIM) in and near the disk using
  the parameterization of \citet{gaensler2008}.  Otherwise the MB, NFW ($C_{\rm v}=3$),
  and DISK line styles are the same as in Figure~\ref{fig:profile}.  Only the corona distributed like an NFW provides a contribution larger than
  the WIM and is clearly ruled out by these data. \textit{Right panel}: We also plot the total DM from the DISK and MB models (see the thick blue and dark curves, respectively).}
\label{fig:dm}
\end{figure*}

The dispersion measure (DM) of pulsars offers a direct probe of the electron
distribution along the sight line towards a background pulsar at a distance $D$
from the Earth:
\begin{equation}
{\rm DM} = \int_0^D \rho_e(r)dr.
\end{equation}
While most known pulsars live in the Milky Way disk, several have been
discovered in the Large and Small Magellanic Clouds (LMC and SMC), with a
distance $D \sim$ 50 and 60 kpc, respectively.

Two recent surveys of pulsars in the LMC and SMC were performed by \citet{crawford2001} and \citet{manchester2006}. They found a total of 21 pulsars, among which we selected 18 pulsars. Five of the selected pulsars are located in the SMC and the rest in the LMC. We discarded 3 pulsars with a ${\rm DM} \sin|b|<25\rm\ cm^{-3}\,pc$ since the
distribution of ${\rm DM}\sin|b|$ of the Galactic pulsars suggests those pulsars
likely lie within the Milky Way disk (\citealp{crawford2001, gaensler2008}). We plot the DM of the remaining pulsars -- which should be associated
with the Magellanic Clouds -- and their errors in Figure~\ref{fig:dm}. For
visual clarity, we randomly assign distances between 50 and 60 kpc for these
pulsars. The pulsar DMs in LMC and SMC range from 40 to $\sim 200\rm\
cm^{-3}\,pc$. Higher DMs for some of the pulsars indicate most likely a
significant contribution from electrons in the LMC or SMC, so the lowest DM will
provide an upper limit on the hot gas distributed between us and the LMC and
SMC.

There are two important potential sources for electrons giving rise to the DM.
The first is the hot gas corona, which we will explore below.  The second is the
warm ionized medium (WIM) in and near the disk.  It has long been suggested that
the distribution of the free electrons in the WIM of the disk follows a planar
distribution (see, e.g., \citep{reynolds1989}):
\begin{equation}
n^{\rm WIM}(z) = n_0^{\rm WIM}\exp\left(-\frac{z}{H_n^{\rm WIM}}\right).
\end{equation}
Here $n_0^{\rm WIM}$ and $H_n^{\rm WIM}$ are the mid-plane WIM electron density
and the scale height, respectively. By fitting a total of 53 sightlines,
\citet{gaensler2008} found $n_0^{\rm WIM}=0.031^{+0.004}_{-0.002}\ \rm
cm^{-3}$ and $H_n^{\rm WIM}=1010^{+40}_{-170}\ \rm pc$; we adopt these
values here.  The dash-dotted line in Figure~\ref{fig:dm} shows the
contribution of this WIM distribution to the DM measure.  We specifically
calculate the contribution along the line of sight toward the LMC.  We see that it is significant, leaving very
little room for additional electrons from a hot halo.

The solid black, dashed red, and dotted blue lines in the left panel of Figure ~\ref{fig:dm} show
the DM coming from our MB, NFW ($C_h = 3$), and DISK models, respectively.  In
each case we calculate the DM within a line-of-sight distance $D$ pointed in the
direction $(l,b)$ via
$D^2 = r^2+D_0^2-2rD_0\cos l \cos b$,
where $D_0 = 8$ kpc is the distance between the Sun and the Galactic Center and
$r$ is the Galactocentric distance.  It is clear that both the MB and DISK
models produce DM contributions that are well below the WIM, such that the total
WIM+DM or WIM+DISK results are consistent with the data.  The $C_h=3$ NFW model,
on the other hand, is clearly inconsistent, especially when the total WIM+NFW
contribution is considered.

Interestingly, the DM from the MB profile continues to rise to large radii. In
principle, if we were confident about the contribution from the WIM, a
dispersion measure from more distant pulsars would provide interesting
constraints such an extended profile.  In the right panel of Figure ~\ref{fig:dm} we show the detailed difference between the total DM of the MB+WIM model (black, solid line) and the DISK+WIM model (blue, dotted line). The two models already predict significant difference at a distance of $\sim$100 kpc. The next brightest distant satellite
(after the SMC and LMC) is the Fornax dwarf at a distance of $\sim 150$ kpc.  To
our knowledge, there are no known pulsars in Fornax \citep{mclaughlin2003},
but further searches would be useful.


\section{Discussion}
\label{sec:discussion}
\subsection{Extended Halo: Parameter Dependence}
\label{subsec:param_dependence}
We have demonstrated that if one considers a hot halo profile with a low density
(high entropy) core, as expected for an adiabatic gas in hydrostatic equilibrium
\citep{maller2004} or for a galaxy-size hot halo in hydrostatic and
thermal equilibrium \citep{sharma2012}, then the Milky Way could in principle
contain the universal baryon fraction within its virial radius.  Specifically,
in our fiducial MB model we considered a Milky Way halo of virial mass $M_{\rm
  v}=10^{12}\rm\ M_{\odot}$ and a hot halo of mass $M_{\rm hot} = 10^{11}$
M$_\odot$, which would make the system baryonically
closed.
The X-ray surface brightness is quite sensitive to the chosen value of the
metallicity of the gas; we have so far used $Z_g=0.3\,Z_{\odot}$. In this
section, we investigate how our MB halo results depend on our specific choice of
virial mass, metallicity, and hot gas mass.

\begin{figure*}[t]
\center
\centerline{\includegraphics[width=.5\textwidth,angle=180]{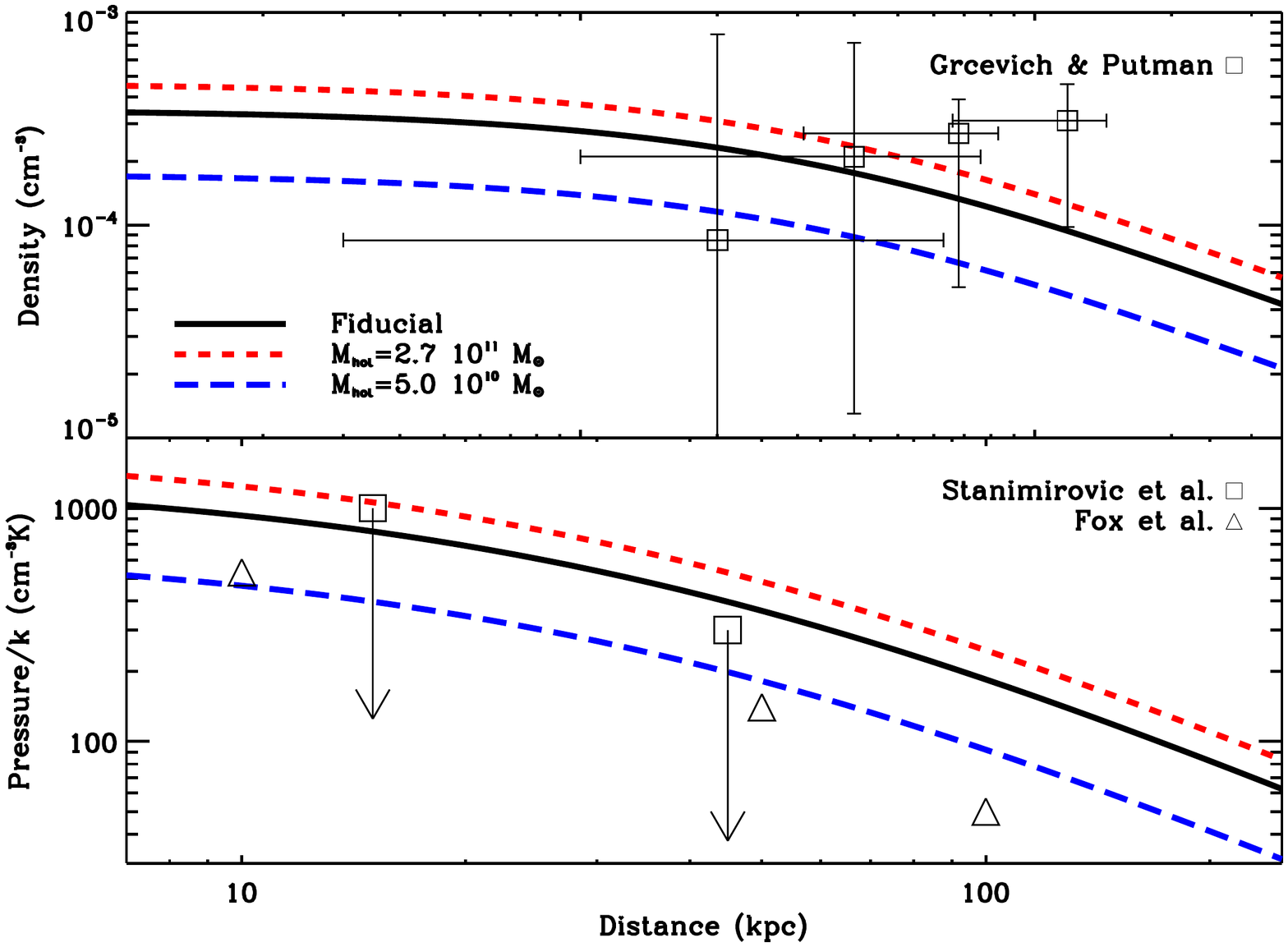}
\includegraphics[width=.5\textwidth,angle=180]{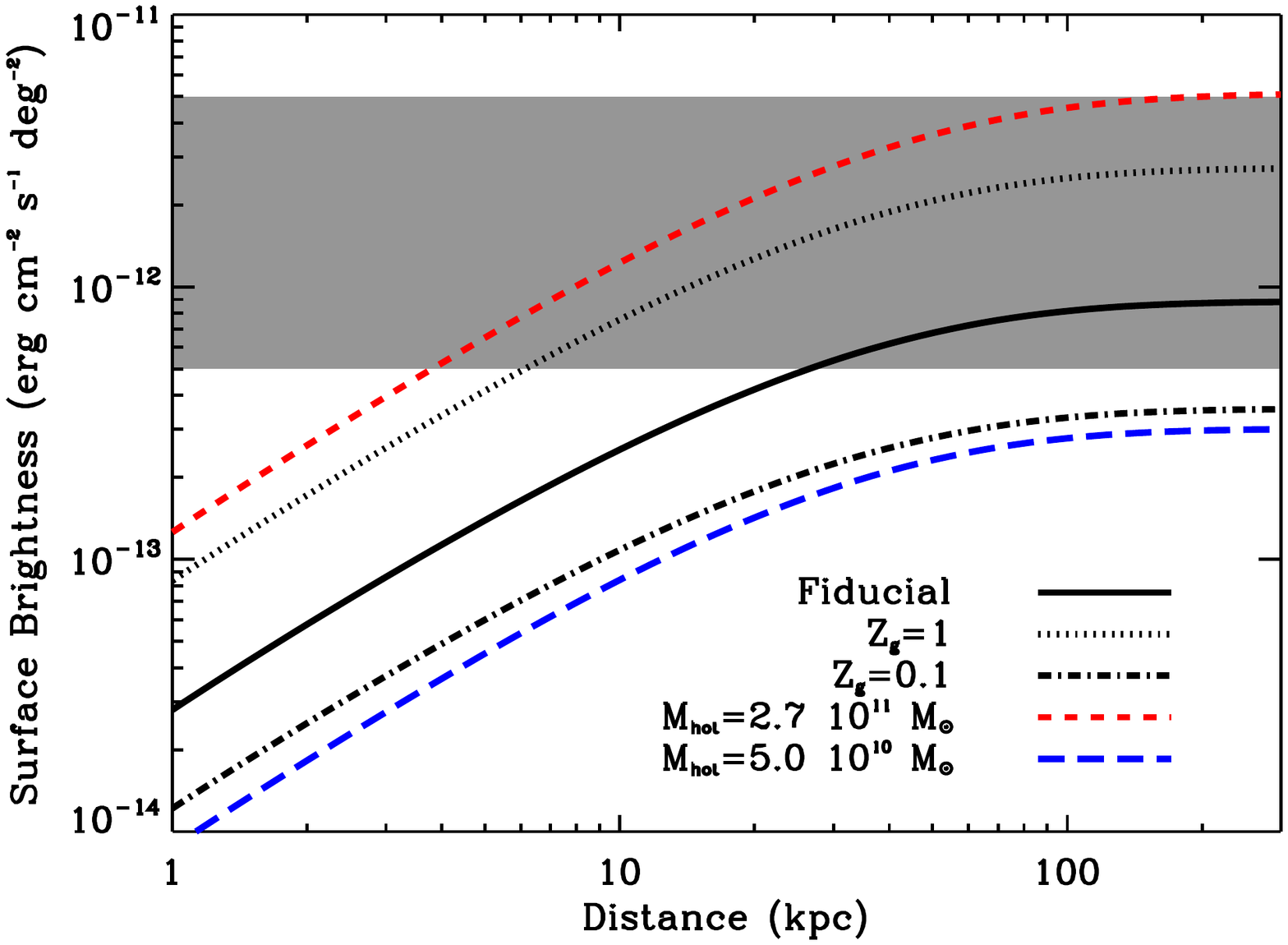}
}
\vskip-.5cm
\caption{\textit{Left panel}: Density (top) and pressure (bottom) distributions for MB
  profiles with $M_{\rm hot} = 2.7$ (red dashed), $1.0$ (solid black,
    fiducial), and $0.5 \times 10^{11}$ M$_\odot$ (blue, long-dashed). \textit{Right
  panel}: The X-ray surface brightness profiles for the same three models are
  shown with the same line types as in the left panel.  In addition, the dotted and dot-dashed
  lines show our fiducial halo with both higher metallicity ($Z_g=1 Z_{\odot}$,
  top) and lower metallicity ($0.1Z_{\odot}$, bottom), respectively.}
\label{fig:profile_mb}
\end{figure*}
 
The red dashed lines in the left panel of Figure~\ref{fig:profile_mb} show
the density (top panel) and pressure (bottom panel) profiles under the
assumption that the Milky Way virial mass is at the upper end of the expected
range, $M_{\rm v}=2\times10^{12}\rm\ M_{\odot}$, and that the system is
baryonically closed ($M_{\rm hot} = 2.7 \times 10^{11}$ M$_{\odot}$). Note that
while the density itself appears to be consistent with the Grcevich \& Putman
determination from HI gas stripping, the implied pressure profile is too high to
explain the properties of gas clouds in the halo.  The solid black lines
  show our fiducial model for reference and the blue, long-dashed lines show a
similar halo that contains only $\sim 2/3$ of its baryons in total, with a hot
halo mass of $M_{\rm hot} = 5 \times 10^{10}$ M$_{\odot}$.  This halo is fairly
consistent with the Fox et al. pressure estimates, but it somewhat too low in
density to explain the lack of HI in Milky Way dwarfs.

The same group of models are presented in the right panel in comparison to the
X-ray surface brightness constraints for our fiducial metallicity assumption
$Z_g = 0.3$. They are all consistent with these data.  The predicted $S_X$
values are also sensitive to metallicity, as metals increase the cooling
efficiency. To investigate the effect of metals, we calculate $S_X$ for two
additional metallicities within our fiducial MB halo: $Z_g=1\,Z_{\odot}$ (the
dotted dark line in Figure~\ref{fig:profile_mb}), and $Z_g=0.1\,Z_{\odot}$ (the
dashed dark line).  Although the predicted $S_X$ is still consistent with
observations for $Z_g=1\,Z_{\odot}$, it is on the high end of the observed
$S_X$. 
 
\section{Summary}
\label{sec:summary}
Whether or not galaxies like the Milky Way host substantial masses of baryons in
hot coronae has a fundamental impact on our understanding of galaxy formation
and evolution. In this paper, we examined the latest X-ray emission and pulsar
dispersion measure data to test for the presence of hot baryons in the distant
halo of our Milky Way. The three models that we used to describe the hot gas
distribution in the halo are: (1) a NFW-type; (2) an extended hot halo motivated
by MB04; and (3) a disk distribution.

We found that for a baryonically closed Milky Way, the hot gas cannot follow an
NFW profile, either for a standard halo concentration ($C_h=12$) or for a low
concentration ($C_h=3$): the NFW profile predicts both too much X-ray emission
and a pulsar dispersion measure that exceeds the latest observational data. The
baryon fraction in such models must be substantially lower ($f_b\sim0.01-0.02$)
to be consistent with data (see also \citealp{anderson2010}).

The other two classes of models we have considered -- extended halos of hot gas
in hydrostatic equilibrium \citep{maller2004} and a hot disk of gas -- can
both be made consistent with existing X-ray emission and pulsar dispersion
measure data. These two models predict very different properties for the hot gas
content of the Milky Way, however: in the DISK model, the hot gas contributes a
negligible amount ($\sim 10^8\,M_{\odot}$) to the Milky Way's baryon budget,
while the MB extended halo model can contain enough hot gas to make the Milky
Way baryonically closed. 

As we were completing this work, Gupta et al. (2012) presented a complementary study of the hot gas content of the Milky Way halo using X-ray spectra of background AGNs. Based on the detected z = 0 absorption lines produced by highly ionized oxygen, O VII and O VIII, and a joint analysis with the Galactic halo emission, Gupta et al. argued for the existence of an extended hot gas around our Galaxy, with a radius of over 100 kpc and a total mass in excess of $10^{10}\,M_{\odot}$. Their conclusion is fully consistent with our results for extended hot gas profiles, and incorporating X-ray absorption data, along with considerations of uncertainties associated with the gas metallicity and ionization mechanism(s), would likely be a fruitful avenue for future extensions of our current work.

Incorporating additional indirect constraints -- i.e., the lack of gas in
\textit{all} Milky Way dwarf spheroidals and the possibility that high velocity
clouds are pressure-confined by a hot ambient medium -- may indicate that a
MB-type model with an extended hot halo is favored. This interpretation should
be testable in the near future, as observations of X-ray emission at low
latitudes ($|b|<20^{\circ}$) can distinguish between the MB and the DISK
models. Such observations ideally should be performed at $90 < l < 270^{\circ}$
to avoid the contamination from the Galactic center. If pulsars can be detected
in distant dwarf spheroidal galaxies with recent star formation -- e.g., Fornax
and Leo I -- dispersion measures for these objects would also be very
constraining for the distribution of hot gas around the Milky Way.

\vspace{0.5cm}
\section*{Acknowledgments}
We thank David Buote and Philip Humphrey for helpful discussions.  TF was partially supported by the National Natural Science Foundation of China under grant No.~ 11243001 and No.~11273021. MB-K acknowledges support from the Southern California Center for Galaxy Evolution, a multi-campus research program funded by the University of California Office of Research.   JSB was partially supported by the Miller Institute for Basic Research in Science during a Visiting Miller Professorship in the Department of Astronomy at the University of California Berkeley. 

{\it Facility:} \facility{Suzaku}; \facility{XMM-Newton}


\end{document}